\documentclass{aa}
\usepackage{times}
\usepackage{psfig}

\def\phibar{\overline\varphi}
\def\phigeo{\varphi_{geo}}
\def\phiarm{\varphi_{arm}}

\def\etot{E_{tot}}

\def\nudot{\dot\nu}
\def\nuddot{\ddot\nu}

\def\smp{\hskip 0.25em}
\def\comptel{{COMPTEL }}

\begin{document}

%%%%%%%%%%%%%%%%%%%%%%%%%%%%%%%%%%%%%%%%%%%%%%%%%%%%%%%%%%%%%%%%%%%%%%%%%%%%%%%%

\thesaurus{06(13.07.2; 08.16.7 PSR B1509-58)}
\title{COMPTEL detection of pulsed $\gamma$-ray emission from \break PSR B1509-58 up to at least 10 MeV}

\author{L.~Kuiper\inst{2} 
\and    W.~Hermsen\inst{2}
\and    J.M.~Krijger\inst{2,7}
\and    K.~Bennett\inst{3}
\and    A.~Carrami\~nana\inst{4}
\and    V.~Sch\"onfelder\inst{1}
\and    M.~Bailes\inst{6}
\and    R.N.~Manchester\inst{5}
       }

% \and    A.~Connors\inst{3}

\institute{    
               Max-Planck Institut f\"ur Extraterrestrische Physik,
               D-8046 Garching, Germany
  \and
               SRON-Utrecht, Sorbonnelaan 2, 
               NL-3584 CA Utrecht, The Netherlands               
  \and
               Astrophysics Division, European Space Research and Technology
               Centre, NL-2200 AG Noordwijk, The Netherlands
  \and
               I.N.A.O.E., Apartado Postal 51 y 216, 
               Puebla 72000, Puebla, M\'exico
  \and
               Australia Telescope National Facility, CSIRO, PO Box 76, 
               Epping NSW 2121, Australia
  \and
               Astrophysics and Supercomputing, Swinburne University of 
               Technology, PO Box 218 Hawthorn, Victoria 3122 Australia
  \and
               Astronomical Institute, Utrecht University, NL-3508 TA Utrecht, 
               The Netherlands
          }

%  \and
%              Space Science Centre, University of New Hampshire,
%              Durham, NH 03824, U.S.A.

\offprints{e-mail: L.M.Kuiper$@$sron.nl}

\date{Received 14 January 1999 / Accepted 8 February 1999}

\maketitle
%\maintitlerunninghead{COMPTEL detection of PSR B1509-58}
% \authorrunninghead{L.Kuiper et al.}

%%%%%%%%%%%%%%%%%%%%%%%%%%%%%%%%%%%%%%%%%%%%%%%%%%%%%%%%%%%%%%%%%%%%%%%%%%%%%%%%

%\psfigurepath{/home/kuiper/PSRB1509-58/PAPER/postscript}

%%%%%%%%%%%%%%%%%%%%%%%%%%%%%%%%%% Main Text %%%%%%%%%%%%%%%%%%%%%%%%%%%%%%%%%%%

  \begin{abstract}

  We report on the first firm detection of pulsed $\gamma$-ray emission from PSR B1509-58 in the 0.75-30 MeV 
  energy range in CGRO COMPTEL data collected over more than 6 years.
  The modulation significance in the 0.75-30 MeV pulse-phase distribution is $5.4\sigma$ and 
  the lightcurve is similar to the lightcurves found earlier between 0.7 and 700 keV: 
  a single broad asymmetric pulse reaching its maximum 0.38 $\pm$ 0.03 in phase after the radio peak, 
  compared to the offset of 0.30 found in the CGRO BATSE soft gamma-ray data, and 0.27 $\pm$ 
  0.01 for RXTE (2-16 keV), compatible with ASCA (0.7-2.2 keV).
  
  Analysis in narrower energy windows shows that the single broad pulse is significantly detected up
  to $\sim 10$ MeV. Above 10 MeV we do detect marginally significant ($2.1\sigma$)
  modulation with an indication for the broad pulse. However, imaging analysis shows the presence of a
  strong 5.6$\sigma$ source at the position of the pulsar. To investigate this further, we have also
  analysed contemporaneous CGRO EGRET data ($>30$ MeV) collected over a nearly 4 year period.
  In the 30-100 MeV energy window, adjacent to the COMPTEL 10-30 MeV range, a $4.4\sigma$
  source can be attributed to PSR B1509-58.
  Timing analysis in this energy window yields an insignificant signal of $1.1\sigma$, but with a shape 
  somewhat similar to that of the COMPTEL 10-30 MeV lightcurve. Combining the two pulse-phase distributions 
  results in a suggestive double-peaked pulsed signal above the background level estimated in the spatial 
  analyses, with one broad peak near phase 0.38 (aligned with the pulse observed at lower energies) and a 
  second narrower peak near phase 0.85, which is absent for energies below 10 MeV. 
  The modulation significance is, however, only $2.3\sigma$ and needs confirmation. 
  
  Spectral analysis based on the excess counts in the broad pulse of the lightcurve shows that extrapolation 
  of the OSSE power-law spectral fit with index -1.68 describes our data well up to 10 MeV. Above 10 MeV the 
  spectrum breaks abruptly. The precise location of the break/bend between 10 and 30 MeV depends on the 
  interpretation of the structure in the lightcurve measured by COMPTEL and EGRET above 10 MeV. 
  
  Such a break in the spectrum of PSR B1509-58 has recently been interpreted in the framework of polar cap 
  models for the explanation of gamma-ray pulsars, as a signature of the exotic photon splitting process in 
  the strong magnetic field of PSR B1509-58.  
  For that interpretation our new spectrum constrains the co-latitude to $\sim 2\degr$, close to the ``classical'' 
  radius of the polar cap. In the case of an outer-gap scenario, our spectrum requires a dominant synchrotron 
  component.


  \end{abstract}

%%%%%%%%%%%%%%%%%%%%%%%%%%%%%%%%%%%%%%%%%%%%%%%%%%%%%%%%%%%%%%%%%%%%%%%%%%%%%%%%

  \section{Introduction}

  PSR B1509-58 was discovered as a 150 ms X-ray pulsar in the {\em Einstein} HRI and IPC
  (0.2-4 keV) data from observations performed in 1979 and 1980 of supernova remnant (SNR) MSH 15-52
  (\cite{seward}). The pulsations and the large period derivative indicated in the X-ray data 
  were soon confirmed at radio-wavelengths (\cite{manchester}), while the derived dispersion measure supports 
  its association with the SNR. The inferred characteristic age is 1570 year and the component of
  the surface magnetic field perpendicular to the spin axis at the magnetic pole is $3.1\times 10^{13}$ Gau\ss, one of 
  the highest among the steadily growing sample of radio-pulsars. Radio-data collected during an 11 yr time span
  showed that the pulsar did not glitch and made a detailed study of its slow-down possible (\cite{kaspi}). The
  measured braking index was $n = 2.837(1)$, close to $n = 3$ expected for a dipole.

  Extensive X-ray studies of PSR B1509-58 and its environment have been performed in the early eighties 
  at soft- and medium X-ray energies using the {\em Einstein} HRI, IPC and SSS (\cite{seward2},1984) 
  and MPC (\cite{weisskopf}) instruments and in the late eighties using the EXOSAT ME and LE instruments
  (\cite{trussoni}). The morphology of MSH 15-52 at X-rays is complex with at the north western rim of the 
  SNR an excess near the H{$\alpha$} nebula RCW 89 and close to the middle of the SNR a clump containing the pulsar
  surrounded by a diffuse synchrotron nebula. The spectrum of the pulsar is hard with a photon power-law
  index of $\sim -1.1\pm0.1$ (\cite{trussoni}). The pulse is broad with a duty cycle of $\sim 50\%$ and rather
  asymmetric: a sharp rise followed by a gradual decline. At soft X-ray energies ($< 2$  keV) there is some 
  indication that the broad pulse is composed of two smaller narrowly separated pulses (EXOSAT LE, 
  \cite{trussoni}; ROSAT PSPC, \cite{becker_thesis}). 

  More recently, the results from ROSAT PSPC/HRI (Greiveldinger et al. 1995, Trussoni et al. 1996, and  
  Brazier \& Becker 1997) and ASCA (\cite{nagase} et al. 1994, \cite{tamura} and \cite{saito}) observations were 
  presented.
  Using the high-spectral resolution of ASCA in combination with imaging (\cite{tamura}) and the high-spatial resolution 
  of the ROSAT HRI (Brazier \& Becker 1997) the morphology of the remnant can be explained by the presence of several 
  components: the pulsar itself, a non-thermal nebula powered by the pulsar with collimated outflow structures (jets) and
  a hot thermal plasma at RCW 89 near the end of the jet. Assuming that the synchrotron nebula surrounding the pulsar can be
  described in terms of a torus and jets similar to the Crab pulsar, the morphology suggests a large angle between pulsar 
  spin axis and line of sight. 

  The detection of pulsed emission at hard X-rays was first reported by \cite{kawai} using Ginga LAC 2-60 keV data.
  They found that the X-ray pulse lags the radio pulse by $0.25\pm0.02$ in phase. The spectrum of the pulsed 
  emission could be represented by a power-law with photon index $-1.3\pm0.05$ confirming its hard nature (\cite{kawai1}).
  Wilson et al. (1993a,1993b) showed that pulsed emission was even detectable in the soft $\gamma$-ray regime using CGRO 
  BATSE data (20-740 keV), confirmed later by \cite{ulmer} (1993) and \cite{matz} using CGRO OSSE data. 
  The OSSE/BATSE pulse phase distribution (lightcurve) showed a phase offset with respect to the radio-pulse of 
  $0.32\pm0.02$, slightly larger than the value obtained from the Ginga data. The OSSE spectrum above 50 keV of the pulsed 
  emission could be described by a power-law with photon index $-1.68\pm0.09$, consistent with the spectral findings, 
  $\alpha = -1.64\pm 0.42$, from the balloon-borne Welcome instrument (94-240 keV) as reported by \cite{gunji}. 
  The spectral measurements by EXOSAT, Ginga, Welcome, OSSE/BATSE suggest a spectral steepening (softening) towards higher
  energies.  
  
  Recently, Rossi XTE absolute timing results on PSR B1509-58 were presented by \cite{rots} showing high resolution pulse 
  profiles for energies between 2 and 128 keV. A comparison of the profiles measured by RXTE (2-16 keV) and BATSE ($>$ 32 
  keV) was made and a 0.03 phase shift of the BATSE lightcurve w.r.t. the RXTE lightcurve, peaking at $0.27\pm 0.01$, was 
  found by the authors. 
  \cite{rots} also performed a pulse-phase resolved spectral analysis using PCA and HEXTE data. The photon indices of the 
  power-law fits to the 2 - 200 keV data in various 0.05 wide phase slices within the pulse are all consistent with one 
  single value of $\alpha = -1.345\pm 0.01$. This value is in line with the value of $-1.3\pm0.05$ found by \cite{kawai}. 
  RXTE spectral results for both the pulsed and unpulsed component are described in detail by \cite{marsden}: the spectrum
  of the pulsed component (radio phase range 0.17 and 0.53) could be represented by a power-law with index $-1.358\pm 0.014$ 
  with no evidence for a spectral break seen up to $\sim 200$ keV, while the index of the unpulsed component was 
  $-2.215\pm 0.005$. 
  
  At high-energy gamma-rays ($> 30$ MeV) Brazier et al. 1994 reported only upper-limits for pulsed emission 
  using CGRO EGRET data from 3 viewing periods (VP 12, 23 and 27; see e.g. Table \ref{obs_table}) during the all-sky survey 
  of CGRO. A study by Fierro (1995) analysing EGRET data from Cycle I-III yielded an interesting $\sim 4\sigma$ source 
  feature for energies above 100 MeV consistent in position with the pulsar. Timing analyses resulted in non-detections for 
  pulsed emission in the 30-100 MeV, $>$ 100 MeV and $>1$ GeV energy windows.

  \begin{table*}[t]
  \caption[]{\label{obs_table} COMPTEL observations used in current study with PSR B1509-58 
  less than $30\degr$ off-axis}
  \begin{flushleft}
  \begin{tabular}{lllccccc}
  \hline\noalign{\smallskip}
  VP \#  & Start Date & End Date   & \multicolumn{2}{c}{Pointing direction} &Off-axis angle
                                                      & Eff.Exposure &EGRET spark- \\
         & TJD$^{\dagger}$ & TJD & l($\degr$)  & b($\degr$)  &($\degr$)  & (3-10 MeV; $10^6\ cm^2s$)         
         &chamber status \\
  \hline\noalign{\smallskip}
  Cycle I &             &             &       &       &      & 18.70  &    \\
  12.0    & 8546.620    & 8560.622    & 310.7 & 22.2  & 25.2 & 8.30   & ON \\
  23.0    & 8700.598    & 8714.534    & 322.2 &  3.0  &  4.6 & 3.74   & ON \\
  27.0    & 8740.573    & 8749.589    & 332.2 &  2.5  & 12.4 & 4.55   & ON \\
  35.0    & 8840.658    & 8845.038    & 335.1 &-25.6  & 28.3 & 0.83   & ON \\
  38.0    & 8861.745    & 8866.192    & 335.1 &-25.6  & 28.3 & 1.28   & ON \\
  \hline\noalign{\smallskip}
  Cycle II&             &             &       &       &      & 11.71  &    \\ 
  208.0   & 9020.627    & 9027.675    & 307.4 & 20.7  & 25.3 & 2.32   & ON \\
  215.0   & 9078.693    & 9083.810    & 311.7 & 22.9  & 25.5 & 4.32   & ON \\
  217.0   & 9089.577    & 9097.596    & 311.7 & 22.9  & 25.5 & $\perp$& ON \\
  232.0   & 9223.612    & 9225.197    & 347.5 &  0.0  & 27.2 & 5.07   & ON \\
  232.5   & 9225.219    & 9237.590    & 347.5 &  0.0  & 27.2 & $\perp$& ON \\
  \hline\noalign{\smallskip}
  Cycle III&            &             &       &       &      & 23.97  &    \\
  314.0   & 9355.682    & 9368.637    & 304.2 & -1.0  & 16.1 & 9.21   & ON \\
  315.0   & 9368.656    & 9375.646    & 304.2 & -1.0  & 16.1 & 5.32   & ON \\
  316.0   & 9375.671    & 9384.606    & 309.5 & 19.4  & 23.1 & 5.39   & ON \\
  336.5   & 9568.601    & 9573.884    & 340.4 &  2.9  & 20.5 & 3.11   & ON \\
  338.0   & 9593.637    & 9595.602    & 345.0 &  2.5  & 24.9 & 0.94   & OFF\\
  \hline\noalign{\smallskip}
  Cycle IV&             &             &       &       &      & 27.01  &    \\
  402.0   & 9643.633    & 9650.588    & 310.3 & -5.0  & 10.7 & 12.71  & ON \\
  402.5   & 9650.614    & 9657.588    & 306.7 & -3.8  & 13.9 & $\perp$& ON \\
  414.3   & 9805.592    & 9811.587    & 347.3 &  0.6  & 27.0 & 2.70   & OFF\\
  423.5   & 9898.619    & 9908.569    & 345.8 & 13.4  & 29.2 & 3.86   & ON \\
  424.0   & 9908.592    & 9923.596    & 312.7 & 19.0  & 21.5 & 7.74   & ON \\
  \hline\noalign{\smallskip}
  Cycle V&             &              &       &       &      & 13.49  &    \\
  516.1   &10160.640    &10163.650    & 341.1 &  5.5  & 21.8 & 1.71   & ON$^{\spadesuit}$ \\
  524.0   &10273.592    &10287.601    & 343.1 & -3.6  & 22.9 & 7.22   & OFF \\
  529.5   &10322.664    &10332.579    & 345.0 &  2.4  & 24.9 & 4.56   & ON$^{\spadesuit}$ \\
  \hline\noalign{\smallskip}
  Cycle VI&             &             &       &       &      & 42.95  &    \\
  619.0   &10574.621    &10582.557    & 319.6 & -1.6  & 0.9  & 6.86   & OFF \\
  619.4   &10588.601    &10596.510    & 319.6 & -1.6  & 0.9  & 6.36   & OFF \\
  619.7   &10603.617    &10609.567    & 319.6 & -1.6  & 0.9  & 5.21   & OFF \\
  632.1   &10728.603    &10755.605    & 307.9 & -7.5  & 13.9 &24.52   & OFF \\
  \hline\noalign{\smallskip}
  \multicolumn{8}{l}{$^{\dagger}$\smp\smp TJD = JD - 2440000.5 = MJD - 40000} \\
  \multicolumn{8}{l}{$^{\spadesuit}$ EGRET in narrow field mode; opening angle FoV is $19\degr$} \\
  \end{tabular}
  \end{flushleft}
  \end{table*}

  The detection of hard pulsed emission below $\sim 700$ keV and the non-detection above 30 MeV makes PSR B1509-58 a 
  very interesting candidate for COMPTEL, the Compton Telescope aboard CGRO and sensitive to photons with energies
  between $\sim 0.75$ and 30 MeV. COMPTEL analysis using viewing period 23 (see Table \ref{obs_table}) data only yielded 
  an interesting $\ga 3\sigma$ detection of pulsed emission in the 0.75-1 MeV energy window with a pulse aligned with the 
  pulse observed by BATSE/OSSE (\cite{hermsen}, \cite{carraminana1}). 
  However, a timing analysis of COMPTEL 0.75-30 MeV data from observations spread over more than 4 years yielded only a 
  marginal detection of pulsed emission at energies below 3 MeV (\cite{carraminana2}).
  Here, we will report on the COMPTEL timing- and spatial analyses using all available data up to and including CGRO 
  Cycle-6 data. Prompted by our results in the highest standard energy interval (10-30 MeV) in the COMPTEL analysis, we
  also (re)analysed all publicly available EGRET data on this source.

%%%%%%%%%%%%%%%%%%%%%%%%%%%%%%%%%%%%%%%%%%%%%%%%%%%%%%%%%%%%%%%%%%%%%%%%%%%%%%%%

  \section{Instrument description and observations}

  \comptel is the imaging Compton Telescope aboard CGRO and is sensitive for $\gamma$-ray
  photons with energies in the 0.75-30 MeV range. Its detection principle relies on a
  two layer interaction: a Compton scatter in one of the 7 upper-detector (D1) modules followed
  by a second interaction in one of the 14 lower-detector (D2) modules. Main measured quantities 
  are the angles ($\chi,\psi$) specifying the direction of the scattered photon (from the interaction loci 
  in D1 and D2) and the energy deposits in the D1/D2 modules where the interactions took place.
  From the last two quantities we can calculate the scatter angle $\phibar$ and the total energy
  deposit $\etot$ (see for a full description \cite{schonfelder}).

  Its energy resolution is 5-10\% FWHM and due to its large field of view of typically 1 
  steradian it is possible to monitor a large part of the sky simultaneously with a position determination 
  accuracy of $\sim 1\degr$. The events are time tagged with an accuracy of 0.125 ms.

  \begin{table*}
  \caption[]{\label{tab_ephemerides} PSR B1509-58 radio-ephemerides}
  \begin{flushleft}
  \begin{tabular}{ccccccccc}
  \hline\noalign{\smallskip}
  \multicolumn{2}{c}{Pulsar position} & \multicolumn{2}{c}{Validity range} & $t^{0}$ & $\nu$ 
  & $\nudot$ & $\nuddot$ & $\phi_{0}$ \\
  $\alpha_{2000}$ & $\delta_{2000}$ & \multicolumn{2}{c}{[MJD]} & [MJD/TDB] & [Hz] &[Hz/s] 
  &[Hz/s$^2$] & \\
  \hline\noalign{\medskip}
  $^{\dagger}$15~13~55.627  &-59~08~9.54    & 48522 & 49956 & 49239  & $6.6324050404788~$ 
  &$-6.75457 \times 10^{-11}$   & $1.96\times10^{-21}$ & $0.93162$ \\
  $^{\ddagger}$15~13~55.620  &-59~08~9.00    & 50114 & 50722 & 50418  & $6.6255346044703~$ 
  &$-6.73467 \times 10^{-11}$   & $1.94\times10^{-21}$ & $0.64393$ \\
  \hline\noalign{\bigskip}
  \multicolumn{9}{l}{$^{\dagger} $\smp Ephemeris has been provided by V. Kaspi (private communication).} \\
  \noalign{\smallskip}
  \multicolumn{9}{l}{$^{\ddagger}$\smp Ephemeris has been been derived from radio timing data using the ATNF Parkes radio telescope.} \\
  \multicolumn{9}{l}{\smp\smp (see also {\em http://www.atnf.csiro.au/research/pulsar/psr/archive/})} \\
  \multicolumn{9}{l}{\smp\smp The dispersion measure used in the calculation of absolute phases was $253.2\ pc\ cm^{-3}$.} \\
  \end{tabular}
  \end{flushleft}
  \end{table*}

  In this study we selected those CGRO Cycle I-VI viewing periods for which the angle between the pointing 
  axis (co-aligned with the COMPTEL/EGRET z-axis) and PSR B1509-58 is less than $30\degr$.
  Details for each individual observation can be found in Table \ref{obs_table}. The last but one column 
  specifies the effective exposure in the 3-10 MeV energy window assuming a $E^{-2}$ dependency of the photon flux. 
  The calculation took into account Earth blocking effects and utilizes the timeline information. 
  The last column indicates the status of the sparkchamber of the EGRET high-energy (30 MeV - 30 GeV) 
  instrument aboard CGRO. 
  EGRET data (with the sparkchamber ON) from the first 4 CGRO observation Cycli 
  have been retrieved from the COMPTON Science Support Center and have subsequently been used in 
  spatial- and timing analyses.

%%%%%%%%%%%%%%%%%%%%%%%%%%%%%%%%%%%%%%%%%%%%%%%%%%%%%%%%%%%%%%%%%%%%%%%%%%%%%%%%

  \section{COMPTEL timing analysis}

  %%%%%%%%%%%%%%%%%%%%%%%%%%%%%%%%%%%%%%%%%%%%%%%%%%%%%%%%%%%%%%%%%%%%%%%%%%%%%%%%

  \subsection{Event selections}

  Prior to the actual timing analysis we have to specify the event selection criteria
  to which the events are subjected. The selection criteria applied here are the same 
  as those in the timing analysis of PSR B1951+32 (see Kuiper 1998a) except for one 
  selection parameter, namely $\phiarm$, the difference angle between the calculated 
  scatter angle $\phibar$ and the geometrical scatter angle $\phigeo$. The last quantity
  can be determined from the known source position (\smp({\sl l,b})=(320\fdg 321,-1\fdg 162)) 
  and the scatter direction angles $(\chi,\psi)$. For a point-source the distribution of $\phiarm$ 
  (i.e. the ARM-distribution) is a narrowly peaked distribution with a maximum near $\phiarm=0$ and 
  a wing for positive $\phiarm$ values due to incompletely absorbed events. 
  The imaging capabilities of COMPTEL rely on this sharp asymmetric distribution of $\phiarm$. 
  The relative contributions of the peak and wing, and the width of the peak are a 
  function of input photon energy.
  This means that instead of fixing $\vert\phiarm\vert$ to a value in the range $2\fdg 5$ to 
  $3\fdg 5$ irrespective of the selected energies, as turned out to be the optimum range from COMPTEL studies 
  on the Crab (\cite{much}) and Vela (\cite{kuiper2}) pulsars, an energy window dependent ARM 
  selection is more appropriate. 
  In this study we have determined {\em a priori} the optimal value of $\vert\phiarm\vert$ for 
  each energy window by estimating the maximum in the Signal-to-Noise vs. $\vert\phiarm\vert$ relation.
  The latter relation can be derived from a 3 dimensional ($\chi,\psi,\phibar$) point source model
  for the energy window involved and the total measured 3d-event distribution in the same energy window, 
  heavily dominated by instrumental background events (90-95\%). 
  The following energy dependent criteria on $\vert\phiarm\vert$ appeared to be appropriate: 
  $3\fdg 5$ for the energy window 0.75-1 MeV and $2\fdg 5$ for the energy windows 1-3, 3-10
  and 10-30 MeV. 
  The fraction of counts from a point-source within the ARM-cut is typically $\sim 60\%$.
  It is also worth mentioning that the ARM cut applied in the timing analysis 
  reduces the number of events handled in the timing analysis to typically 10\% of the 
  number of events available for the imaging or spatial analysis, in which the full 3d-dataspace is employed. 

  For the 10-30 MeV interval we have departed from the ``standard'' Time of Flight (TOF) and Pulse Shape 
  Discrimination (PSD) windows (see for a description of these event parameters \cite{schonfelder}) 
  of 113-130 and 0-110 respectively, normally applied in the timing analysis 
  (see e.g. Kuiper 1998a) and have used the optimum TOF and PSD windows derived by \cite{collmar} in their
  study on optimum parameter cuts using the Crab pulsar/nebula signature in the COMPTEL event space.

  %%%%%%%%%%%%%%%%%%%%%%%%%%%%%%%%%%%%%%%%%%%%%%%%%%%%%%%%%%%%%%%%%%%%%%%%%%%%%%%%

  \subsection{Pulse phase folding}

  Once the event selection criteria were settled we proceeded as follows: the arrival times (recorded with an 
  intrinsic resolution of 0.125 ms) of the events passing through our selection filters are converted to 
  arrival times at the Solar System Barycentre (SSB) using the known instantaneous spacecraft position, the 
  source position and the solar system ephemeris (JPL DE200 Solar System Ephemeris). 
  The pulse phase $\phi$ is calculated from the following timing model:

  \begin{equation}
      \phi = \nu \cdot \Delta t + 1/2 \cdot \nudot \cdot \Delta t^2 +
      1/6 \cdot \nuddot \cdot\Delta t^3 - \phi_0
      \label{eq:timing}
  \end{equation}

  \noindent In this formula $\Delta t$ is given by $\Delta t = t^{e} - t^{0}$ with $t^{e}$
  the event SSB arrival time and $t^{0}$ the reference epoch.
  The values employed here for $t^{0},\nu,\nudot,\nuddot,\phi_0$ are given in Table \ref{tab_ephemerides}.
  The RMS error of the timing models listed in Table \ref{tab_ephemerides} is typically 10 milli-periods or
  1.5 ms, sufficiently accurate to keep coherency and allowing pulse phase folding over long time spans indicated
  by the validity range.

  %%%%%%%%%%%%%%%%%%%%%%%%%%%%%%%%%%%%%%%%%%%%%%%%%%%%%%%%%%%%%%%%%%%%%%%%%%%%%%%%
  
  \subsection{Pulse profiles in the 0.75-30 MeV energy range}

\begin{figure}[h]
     \vspace{-0.25truecm}
     \hbox{
           \hspace{0.5cm} 
           \psfig{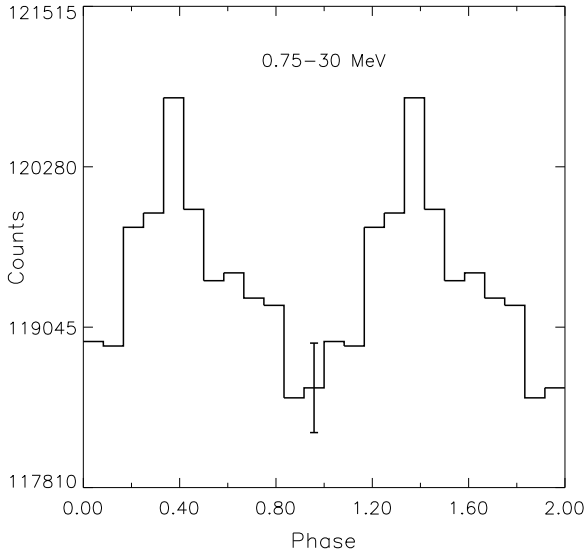}
          }
     \caption[]{Radio-aligned (radio pulse at phase 0) 0.75-30 MeV \comptel lightcurve of PSR B1509-58. 
     A double cycle is shown
     for clarity. Data from Cycle I-VI observations (see Table {\ref{obs_table}}) have been used. The modulation
     significance is $5.4\sigma$ adopting a $Z_2^2$-test. A typical error bar is indicated.}
     \label{fig_comptel_lc_integral}
  \end{figure}

  The pulse phase distribution resulting from phase-folding COMPTEL Cycle I-VI 0.75-30 MeV data is shown in Fig. 
  \ref{fig_comptel_lc_integral}. The modulation significance of the unbinned sample of pulse-phases is $5.4\sigma$
  employing a $Z_n^2$-test (\cite{buccheri}) with 2 harmonics. This is the first {\em firm} detection of pulsed
  emission above 0.75 MeV from PSR B1509-58. The pulse is roughly aligned with the pulse observed by OSSE/BATSE 
  (\cite{ulmer} 1993) and peaks at phase $0.38\pm0.03$ (obtained from a Gaussian + background fit).

\begin{figure}[t]
     \vspace{-0.85truecm}
     \hbox{
           \hspace{0.35truecm} 
           \psfig{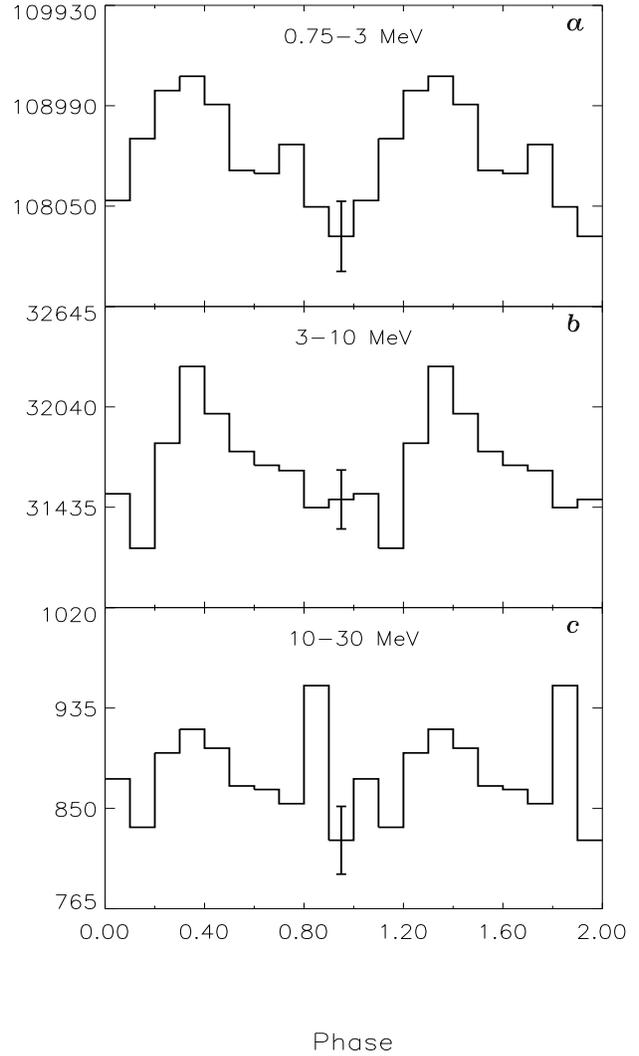}
          }
     \vspace{0.75truecm}
     \caption[]{Radio-aligned \comptel Cycle I-VI lightcurves of PSR B1509-58 for 3 different energy windows:
     0.75-3 MeV ({\bf a}), 3-10 MeV ({\bf b}) and 10-30 MeV ({\bf c}). Double cycli are shown for clarity, while
     typical error bars are indicated. The modulation significances adopting a $Z_2^2$-test are $3.7\sigma, 4.0\sigma$
     and $2.1\sigma$ for the energy windows 0.75-3 MeV, 3-10 MeV and 10-30 MeV respectively. This proves that 
     pulsed emission is detected at least up to 10 MeV.}
     \label{fig_comptel_lc_collage}
  \end{figure}

  We have split the integral energy window of 0.75-30 MeV up into 3 smaller energy windows, 0.75-3 MeV, 3-10 MeV and 
  10-30 MeV and performed similar timing analyses. The modulation significances ($Z_2^2$-test) found for the 3 energy 
  windows are $3.7\sigma, 4.0\sigma$ and $2.1\sigma$ respectively, proving the detection of pulsed emission up to at least 
  10 MeV. 
  The lightcurves are shown in Fig.{\ref{fig_comptel_lc_collage}}. The 10-30 MeV lightcurve, having at face value a 
  non-significant modulation, shows an indication for an enhancement in the phase range containing the pulse at lower 
  energies. However, a narrower pulse might be visible near phase 0.85, which is absent at lower energies. 
  A more detailed discussion on the 10-30 MeV lightcurve will be given once the spatial analysis, enabling the 
  measurement of the {\em total} flux from a source, has been introduced.

  %%%%%%%%%%%%%%%%%%%%%%%%%%%%%%%%%%%%%%%%%%%%%%%%%%%%%%%%%%%%%%%%%%%%%%%%%%%%%%%%

  \subsection{Pulse profiles from soft X-rays to medium energy $\gamma$-rays}

  Rots et al. (1998) studied the pulse shape of PSR B1509-58 as a function of energy at medium and hard X-ray energies.
  He found that the RXTE 2-16 keV pulse peak lags the radio peak by $0.27\pm 0.01$ in phase. This value is consistent
  with the lag of $0.25\pm 0.02$ found in Ginga 2-60 keV data by \cite{kawai}. {\cite{ulmer}} (1993) found at
  soft $\gamma$-ray energies a phase lag of $0.32\pm0.02$ in CGRO BATSE and OSSE data. The difference of $0.05\pm 0.022$ 
  ($2.3\sigma$ effect) between these values was considered troublesome by {\cite{rots}}. 
  It could not be attributed to a CGRO clock absolute timing uncertainty and this triggered {\cite{rots}} to reprocess 5 years 
  of BATSE data yielding now a slightly smaller phase lag of 0.30 consistent both with Ulmer's previous estimate and also with 
  the RXTE estimate.

  \begin{figure}[h]
     \vspace{-1.0truecm}
     \hbox{\hspace{0.25truecm}
           \psfig{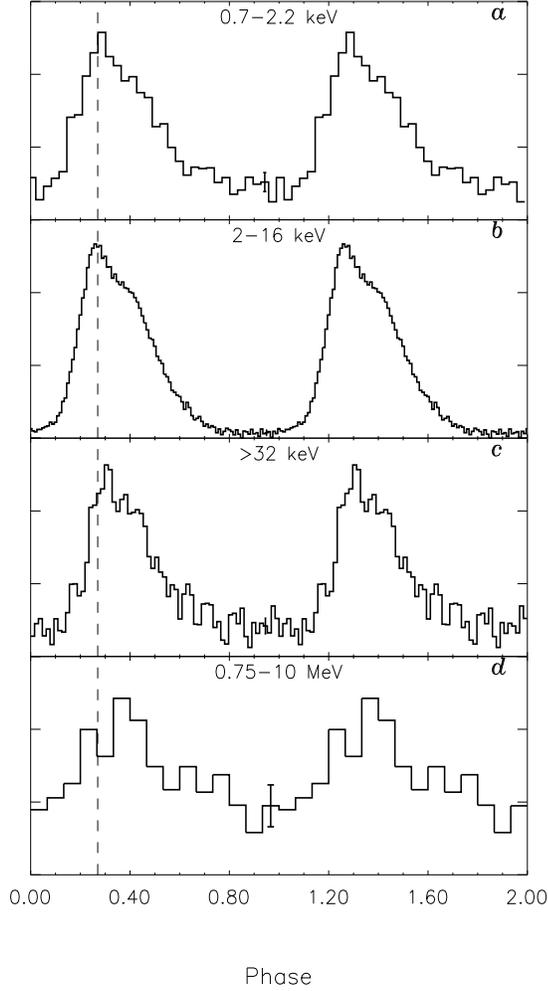}
          }
     \vspace{0.35truecm}
     \caption[]{Radio-aligned lightcurves of PSR B1509-58 from soft X-rays to medium energy
      $\gamma$-rays: {\bf a}) ASCA 0.7-2.2 keV (\cite{saito}; private communication), {\bf b}) RXTE 2-16 keV (Rots et al. 1998),
      {\bf c}) BATSE $>$ 32 keV (Rots et al. 1998) and {\bf d}) COMPTEL 0.75-10 MeV. The dashed vertical line indicates the phase
      lag of 0.27 found in the RXTE lightcurves. Note that the ``center of mass'' of the pulse appears to shift towards 
      higher pulse phases for higher energies, with
      the maximum of the pulse at MeV-energies (COMPTEL) coinciding in phase with the ``shoulder'' visible in the 
      pulse shape at X-ray energies. Typical $1\sigma$ error bars are indicated in each figure at phase $0.95$. 
      }
     \label{fig:lc_collage_instr}
  \end{figure} 

  The remaining (insignificant) difference of $0.03\pm0.022$ ($=5$ ms) still seems to be too large to be explained by
  CGRO/RXTE clock uncertainties and could be due to intrinsic pulsar emission properties. {\cite{rots}} also investigated
  the effect of the pulsar dispersion measure on the phase offset of the radio- and RXTE X-ray pulse.
  Using the most recent value of the dispersion measure of PSR B1509-58 of $255.3\pm0.3\ pc\ cm^{-3}$ instead of the widely 
  used value of $253.2\pm1.9\ pc\ cm^{-3}$ results in an offset of $0.29\pm 0.01$, indicating that uncertainties in the 
  dispersion measure can result in uncertainties in the radio/X-ray pulse phase lag of $\sim 0.02$.
 
  The RXTE 2-16 keV profile consists of a broad asymmetric pulse: a rapidly rising pulse reaching its maximum near phase 0.27
  followed by a more gradual decline with a ``shoulder'' near phase 0.40. This profile is shown in Fig.{\ref{fig:lc_collage_instr}}b
  along with the X-ray/soft $\gamma$-ray profiles measured by different instruments in different energy windows. 
  It is clear from this collage that the ASCA 0.7-2.2 keV soft X-ray profile is in detail very similar to the RXTE 2-16 keV hard 
  X-ray profile. The BATSE $>$ 32 keV soft $\gamma$-ray profile also resembles the RXTE profile except for an apparent overall 
  offset of $0.03$ with respect to the RXTE profile. The COMPTEL 0.75-10 MeV lightcurve on the other hand reaches its maximum near 
  0.38 which coincides in phase with the shoulder clearly visible in the RXTE lightcurve. Whether this apparent offset has a 
  statistical origin (cf. the typical error bar in the COMPTEL 0.75-10 MeV lightcurve which has a mean level of $\sim 93000$ counts
  in this 15 bin lightcurve) or is due to an intrinsic property of the pulsar's high-energy emission is difficult to decide. 
  The intrinsic timing-resolution of 0.125 ms of CGRO/COMPTEL is sufficiently accurate to allow the lightcurve to be binned in 
  several hundreds of bins and can {\em not} be responsible for the offset. Also, BATSE and COMPTEL use the same CGRO clock.

  In order to study the difference in morphology between the COMPTEL 0.75-10 MeV and RXTE 2-16 keV lightcurves we have fitted 
  the RXTE 2-16 keV profile in terms of a background and two Gaussians (7 free parameters). This resulted in a narrow component 
  peaking at phase $0.250\pm 0.008$ with width $0.056\pm0.008$ and a broader component at $0.386\pm 0.012$ and width 
  $0.129\pm0.006$. The first narrow pulse accounts for $25.7\pm4.3\%$ of the total pulsed emission.
  A similar fit has been performed on COMPTEL 0.75-10 MeV data, but now with positions and widths fixed to the values found in 
  the RXTE 2-16 keV fit (3 free parameters). In this case the first narrow pulse can account for only $13\pm18\%$ of the total 
  pulsed emission, consistent with being absent, and the profile can satisfactorily be described by just the broad second
  pulse near 0.39. This strongly suggests that the pulse shape changes from soft X-rays to medium energy $\gamma$-rays. 

  %%%%%%%%%%%%%%%%%%%%%%%%%%%%%%%%%%%%%%%%%%%%%%%%%%%%%%%%%%%%%%%%%%%%%%%%%%%%%%%%

  \subsection{Pulsed 0.75-30 MeV fluxes from the excess counts in the lightcurves}

  Based on the RXTE 2-16 keV lightcurve we have defined a ``pulsed'' and an ``unpulsed'' phase interval in the
  pulse phase distribution: the pulsed interval extends from phase 0.15 to 0.65 and the unpulsed (background) interval
  from 0.65 to 1.15. This break-down is such that for a pulse shape as measured by RXTE $90.4\%$ of the pulse is located in the 
  ``pulsed'' interval.

  Applying this definition to the COMPTEL data we can determine the pulsed excess counts in various energy slices by estimating
  the underlying background as the averaged level in the unpulsed part of the lightcurve. We derived these pulsed excess counts 
  for the 0.75-3, 3-10 and 10-30 MeV energy windows and converted these to pulsed flux values taking into account efficiency
  correction factors due to the applied ARM cuts in the timing-analysis (see Table \ref{tab_pulsed_fluxes}). 
  
  \begin{table}
  \caption[]{\label{tab_pulsed_fluxes} PSR B1509-58 pulsed fluxes derived from the timing analysis with the ``pulsed'' interval
  defined to be phases 0.15-0.65.}
  \begin{flushleft}
  \begin{tabular}{lc}
  \hline\noalign{\smallskip}
  $E$-window & Flux \\
    & $ph / cm^2\cdot s\cdot MeV$ \\
  \hline\noalign{\medskip}
  $0.75-3$  MeV & $(3.69\pm0.73)\times 10^{-5}$ \\
  $3.0-10$  MeV & $(4.52\pm0.77)\times 10^{-6}$ \\
  $10.0-30$ MeV & $(1.21\pm0.85)\times 10^{-7}$ \\
  \hline\noalign{\medskip}
  \end{tabular}
  \end{flushleft}
  \end{table}

  The weak 10-30 MeV flux value should be treated with care because we do not detect significant modulation ($2.1\sigma$). Moreover,
  the lightcurve shows indications for a second pulse in the ``unpulsed''(background) phase interval. If this pulse is genuine, 
  then the true flux is underestimated (see next section).

%%%%%%%%%%%%%%%%%%%%%%%%%%%%%%%%%%%%%%%%%%%%%%%%%%%%%%%%%%%%%%%%%%%%%%%%%%%%%%%%

  \section{COMPTEL spatial analysis}

  \subsection{Analysis method}

  The measured event parameters {\small ($\chi,\psi,\phibar,\etot$)} constitute a 4-dimensional
  data space, in which we have to search for a "source signature". In practice the
  dimension of the data space is lowered by integrating along the $\etot$ direction between user 
  selected boundaries. 
  The event distribution of a point source (the Point Spread Function, PSF) in this 
  reduced 3-d data space {\small $(\chi,\psi,\phibar)$} is concentrated in a cone-shaped 
  structure with its apex at the source position ($\chi_0,\psi_0$). In the spatial or 
  imaging analysis we proceed as follows: We generate a background model from the sparsely filled 
  event dataspace (3d) through sophisticated smoothing techniques (see e.g. {\cite{bloemen2}}). 
  Because the measured events in the 3d-dataspace are primarily internally generated background events 
  ($\sim 90-95\%$) this background model represents a good approximation of the genuine instrumental background.
  The search for point sources in the measured 3d-dataspace is accomplished by a maximum likelihood ratio (MLR) test
  at scan positions in the selected sky field. In the null hypothesis ${\cal{H}}_0$ the measured events are described 
  in terms of a background model alone, while in the alternative hypothesis ${\cal{H}}_1$ the data are described in
  terms of a background model and a point source at a given scan position. From the parameter optimizations under
  ${\cal{H}}_1$ we can derive the source flux. From the optimized likelihoods under ${\cal{H}}_1$ and
  ${\cal{H}}_0$ we can determine the maximum likelihood ratio $\lambda$ at each scan position, giving us information 
  on the detection significance of a source. For more detailed information see {\cite{kuiper1}}.
  
  \subsubsection{Imaging results for energies $<$ 10 MeV}

\begin{figure}[t]
     \hbox{
           \hspace{0.25cm} 
           \psfig{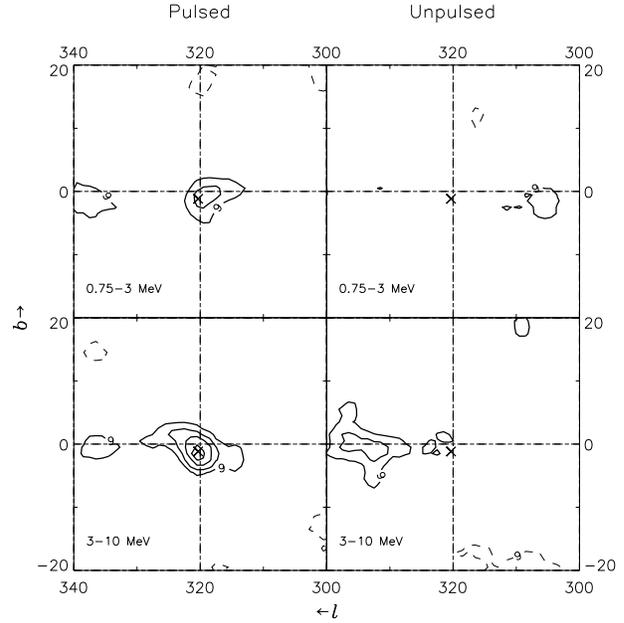}
          }
     \vspace{0.1truecm}
     \caption[]{COMPTEL ``pulsed'' (phases 0.15 - 0.65) and ``unpulsed'' (0.65 - 1.15) MLR maps for energies 
                below 10 MeV.
                The upper panel shows the results for the 0.75-3 MeV energy window, the lower
                panel the 3-10 MeV analogon. The left side shows the pulsed maps and the right 
                side the unpulsed maps. The solid contours start at the $3\sigma$-significance level
                ($\lambda=9$,  1 degree of freedom) in steps of $1\sigma$. Significant emission at the pulsar
                position ({\bf $\times$} symbol) is only observed in the ``pulsed'' maps.}
     \label{fig:comptel_maps_lt_teen}
  \end{figure}

  For consistency purposes we have produced MLR maps for the same energy windows (0.75-3 and 3-10 MeV) as used in the
  timing analysis, and also made selections on pulse phase, the ``pulsed'' and ``unpulsed'' windows introduced in Sect. 
  3.5, while we also considered the total emission, the sum of both. In Fig. \ref{fig:comptel_maps_lt_teen} the MLR 
  images are shown for the ``pulsed'' and ``unpulsed'' phase intervals in the energy windows 0.75-3 and 3-10 MeV. 
  The contours start at an equivalent of $3\sigma$ in steps of $1\sigma$ for 1 degree of freedom (i.e. at $\lambda=9, 
  16, 25, 36$ etc.), representative if the source position is {\em a priori} known. 

  From these maps it is evident that significant emission at the pulsar position (indicated by a ``{\boldmath $\times$}'' mark)
  is only observed in the ``pulsed'' maps: $4.5\sigma$ in the 0.75-3 MeV energy window and $8\sigma$ in the 3-10 MeV 
  window. 
  From the maximum likelihood fits we can also obtain estimates for the source flux. However, in the maps shown in 
  Fig. \ref{fig:comptel_maps_lt_teen} the structured galactic diffuse emission is not included in the background model, 
  and may contribute to the source flux.
  In addition, any pulsar/nebula DC-emission will contribute equally to the ``pulsed'' and ``unpulsed'' maps. 
  Since $\lambda$ is not zero at the source position in the ``unpulsed'' maps, we subtracted the measured fluxes 
  (counts) in the ``unpulsed'' maps from those in the ``pulsed'' maps in order to obtain independent estimates of the 
  ``pulsed'' fluxes. 
  For the 0.75-3 MeV interval we found $7202\pm 1386$ counts correlating with a point source at the pulsar position in 
  the ``pulsed'' interval and $2264\pm 1382$ in the ``unpulsed'' interval. The difference of $4938\pm 1960$ is 
  consistent with the pulsed excess counts in the timing analysis. The number of counts found in the ``unpulsed'' window
  can be fully attributed to contributions of galactic diffuse origin, as has been verified in a simulation of the 
  galactic diffuse emission using the model components and scale factors found by \cite{bloemen}. 
  This means that there is no room for any significant pulsar or nebula DC-component in this energy window.

  For the 3-10 MeV energy interval we obtained a similar picture. In this case we find $5233\pm 593$ counts correlating
  with a point source at the pulsar position in the ``pulsed'' map and $2755\pm 590$ in the ``unpulsed'' map.
  This leaves $2478\pm 840$ counts for the genuinely pulsed emission, consistent with the timing analysis results.  
  The ``unpulsed'' value can again be explained with being of galactic diffuse origin.

  In conclusion, the fluxes derived from the spatial analysis for energies below 10 MeV are consistent with those derived from
  the timing analysis and there is no evidence for the detection of DC-emission from the pulsar or its nebula.  

  %%%%%%%%%%%%%%%%%%%%%%%%%%%%%%%%%%%%%%%%%%%%%%%%%%%%%%%%%%%%%%%%%%%%%%%%%%%%%%%%

  \subsubsection{Imaging results for energies $>$ 10 MeV}

  In the 10-30 MeV energy window we did not detect a significant ($2.1\sigma$) modulation in the light curve, but 
  an enhancement is visible in the pulse phase interval in which the pulse is concentrated at lower energies, and in the ``unpulsed''
  interval a high bin shows up near phase 0.85. A MLR image (total) in this energy window yielded somewhat surprisingly a strong source
  feature ($\sim 6\sigma$) consistent in position with the pulsar, by simply fitting the source contribution above the instrumental
  background model. 
 
  It should be noted here that the 10-30 MeV energy window is COMPTEL's ``cleanest'' window not polluted by time-varying instrumental 
  background lines contrary to the energy intervals below 10 MeV. Estimates for the galactic diffuse emission in this energy window 
  can be considered reliable (e.g. \cite{strong}, Bloemen et al. 1999) and are consistent with extrapolations towards lower energies of
  EGRET $> 30$ MeV measurements. 

  \begin{figure}[t]
     \hbox{\hspace{0.25cm} 
           \psfig{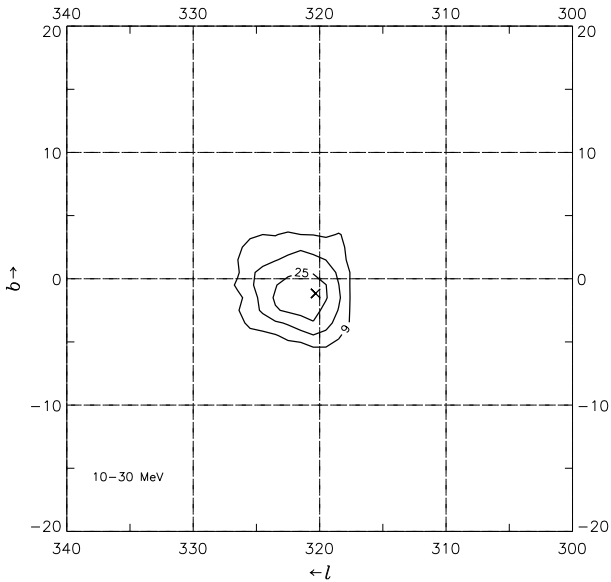}
          }
     \caption[]{COMPTEL ``total'' MLR map for the energy window 10-30 MeV, presenting
                the source detection significance above the instrumental and galactic
                diffuse background. A $5.6\sigma$ source is detected at the pulsar position.   
                The solid contours start at the $3\sigma$-significance level
                ($\lambda=9$,  1 degree of freedom) in steps of $1\sigma$. The pulsar
                position is marked by the {\bf $\times$} symbol.}
     \label{fig:comptel_aboveten}
  \end{figure}

  When we now include  also the total galactic diffuse background model in the total background, the source remains (see
  Fig. {\ref{fig:comptel_aboveten}}). At the pulsar position a detection significance of $5.6\sigma$ is reached. A break-down
  into ``pulsed'' and ``unpulsed'' maps yields sources at the pulsar position with statistically equal fluxes in both maps. 
  This can mean that we either detected DC-emission from the pulsar and/or its surrounding synchrotron nebula or 
  emission from a nearby unrelated source. But, in these cases we would expect to see emission below 10 MeV
  where COMPTEL is more sensitive. Another possibility is that we actually detected pulsed emission but with a different pulse profile. 
  
  We investigated the latter possibility further. In this respect it is instructive to go back to the 10-30 MeV lightcurve and 
  superpose the background level as determined in the spatial analysis from the total number of counts correlating with a source at the 
  pulsar position above the instrumental and galactic diffuse background. 
  The lightcurve with this spatially determined background level (and its $1\sigma$-errors)
  is shown in Fig. \ref{fig:comptel_egret}a. The existence of the main pulse, definitely detected below 10 MeV, is again very 
  suggestive for energies above 10 MeV. The number of excess counts in the ``pulsed''
  interval (0.15-0.65) above of this spatially determined background level is $279\pm57$ counts, resulting in a flux of 
  $(3.37\pm0.70)\cdot 10^{-7}\ ph/cm^2\cdot s\cdot MeV$. In the ``unpulsed'' region we find then a flux of 
  $(2.16\pm0.70)\cdot 10^{-7}\ ph/cm^2\cdot s\cdot MeV$, predominantly due to the excess near phase 0.85. This $\sim 3\sigma$ feature
  above the spatially determined background, only shows up above 10 MeV and,
  if genuine, would indicate a new pulse component with a completely different spectral behaviour
  from that of the main pulse. This peculiar behaviour of the pulsed emission above 10 MeV in the COMPTEL energy range motivated us to 
  analyze also the contemporaneous $>30$ MeV EGRET data.

  %%%%%%%%%%%%%%%%%%%%%%%%%%%%%%%%%%%%%%%%%%%%%%%%%%%%%%%%%%%%%%%%%%%%%%%%%%%%%%%%

  \section{Analysis of EGRET data}

  No significant (pulsed) emission above 30 MeV has been reported from EGRET data, but, two 
  papers (\cite{brazier1}; \cite{fierro}) report on weak (irregular) source features ($\sim 3\sigma$) near PSR B1509-58 for 
  energies above 100 MeV. It is worth mentioning that in the first paper the analysis was constrained to data from CGRO Cycle I,
  namely VP's 12, 23 and 27 (cf. Table 1), while in the second paper data from Cycles I to III were analyzed.  
  Here, we reanalyzed all available Cycle I-IV EGRET data, the maximal exposure on the source. The observations used are those
  in Table 1 in which the spark chamber high-voltage was enabled (see last column Table 1; the data of Cycle V do not contribute since 
  EGRET operated in its narrow field mode with an effective FoV opening angle of $19\degr$). 
  Standard event selections were applied, however, the requirement of a minimum energy deposit of 6.5 MeV in the TASC was abandoned
  %no threshold selection of 6.5 MeV on the TASC was used 
  (for a description of EGRET see \cite{egret}). 
  This selection criterion is mainly effective for background suppression for measured total energies above 100 MeV.
  We verified using the Crab pulsar that for energies below 100 MeV the detection of the pulsed signal significantly increases when the 
  TASC threshold of 6.5 MeV is ignored.
  In order to suppress Earth albedo $\gamma$-rays, standard Earth horizon angle cuts were applied for the differential
  energy ranges 30-50 MeV, 50-70 MeV and 70-100 MeV, which are roughly equivalent to a 3$\sigma$ cut.  
  For consistency, the same selections were applied in the spatial and timing analyses.

  \subsection{Spatial analysis}
  
  The imaging analysis of EGRET data is similar to that of COMPTEL data with the main difference that the EGRET analysis is performed
  in a 2-d dataspace. After reformatting the EGRET data and response for import in the COMPTEL Analysis and Software System COMPASS 
  (\cite{devries}) we can use the same analysis programmes. One important difference is that the EGRET data are almost free of 
  instrumental background.
  This means that the point source emission is searched for above of the dominating galactic and extragalactic $\gamma$-ray 
  backgrounds only.
  The spatial structure of the first component, predominantly due to the interaction of cosmic rays and interstellar Hydrogen, is
  approximated by the measured total column density of atomic Hydrogen H\,{\sc i} and molecular Hydrogen as traced by the CO-molecule, 
  identical to our analysis of COMPTEL data. The latter component can simply be described by an isotropic model.

\begin{figure}[t]
     \hbox{\hspace{0.25cm} 
           \psfig{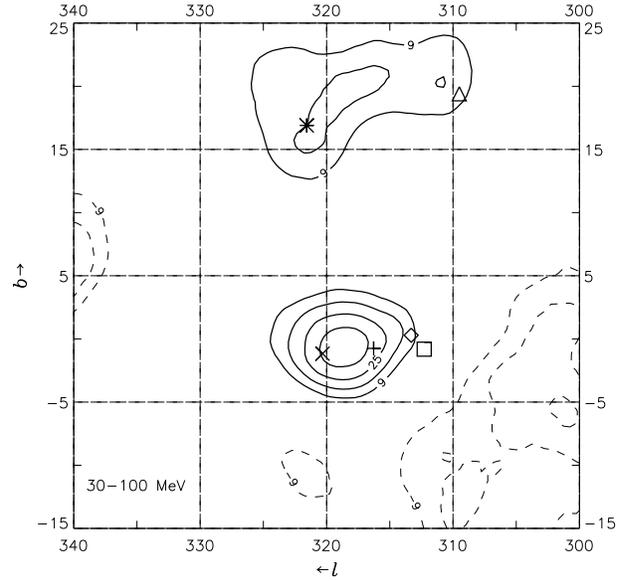}
          }
     \caption[]{EGRET ``total'' MLR map for the energy window 30-100 MeV, showing the likelihood of a source detection
               above of the galactic and extragalactic diffuse backgrounds. Contours of $\lambda=$ 9, 16, 25 (= 3, 4, 5 $\sigma$);
               Solid contours positive values, broken ones negative values. Several identified 
               and unidentified sources are indicated by different symbols: PSR B1509-58 ({\bf $\times$}), Cen-A ({\bf $\triangle$}), 
               and the unidentified EGRET sources 2EG J1412-6211 ({\bf $\sq$}), 2EG J1443-6040 ({\bf $+$}), 2EGS J1418-6049 
               ({\bf $\diamond$}) and 2EGS J1429-4224 ({\bf $\ast$}) (see \cite{thompson1} and \cite{thompson2}).  
               A $6.7\sigma$ source feature near $(l,b)=(320,0)$ shows up and is likely composed of the contributions of PSR B1509-58
               and the unidentified EGRET source 2EG J1443-6040.}
     \label{fig:egret_map1}
  \end{figure}

  The MLR map for the 30-100 MeV energy window, adjacent to the COMPTEL 10-30 MeV window, combining all available Cycle I-IV data is
  presented in Fig. \ref{fig:egret_map1}.  A strong $6.7\sigma$ excess shows up near PSR B1509-58. Apart from the pulsar a few other 
  identified and unidentified EGRET sources seem to be visible in the map (see figure caption).
  
\begin{figure}[t]
     \hbox{\hspace{0.25cm} 
           \psfig{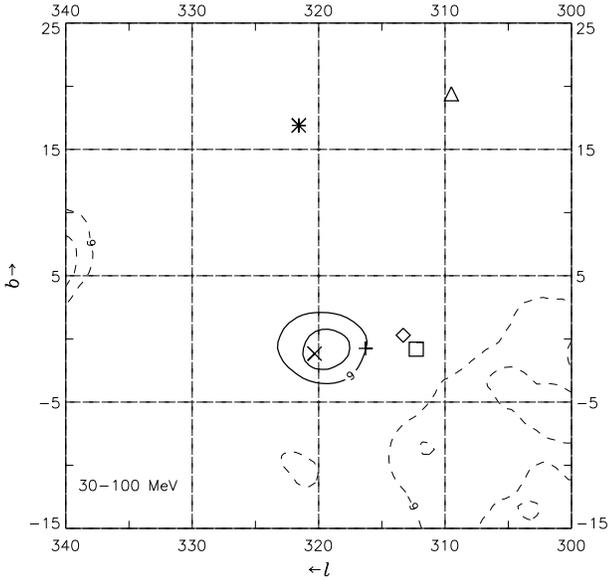}
          }
     \caption[]{EGRET MLR map for the energy window 30-100 MeV, as in Fig. \ref{fig:egret_map1}. 
                The contributions from 2EG J1443-6040, Cen-A and 2EGS J1429-4224 are subtracted. 
                At the position of PSR B1509-58 ({\bf $\times$}) a significance for a source detection of $4.4\sigma$ remains.}
     \label{fig:egret_map2}
  \end{figure}

  The most likely sources responsible for the excess near $(l,b)=(320,0)$ are PSR B1509-58 and 2EG J1443-6040 at 
  $(l,b)=(316.28,-0.75)$. Fitting both sources simultaneously above of the galactic and extragalactic backgrounds, together with 
  emissions from Cen-A and 2EGS J1429-4224 yields a $4.4\sigma$ excess at the position of PSR B1509-58 with a flux of 
  $(3.3\pm1.0)\cdot 10^{-8}\ ph/cm^2\cdot s \cdot MeV$. This is shown in the MLR map of Fig. \ref{fig:egret_map2} in which the other
  three fitted sources and background emissions are ``subtracted''. The detection significance of the source correlating with 
  2EG J1443-6040 is $2.6\sigma$ and has a flux of $(1.9\pm1.0)\cdot 10^{-8}\ ph/cm^2\cdot s \cdot MeV$. 
  It should be noted that this unidentified EGRET source was only detectable in VP's 12 and 27 and not during
  observations performed later on, indicating its likely transient behaviour. In fact, it does not appear anymore in the third EGRET
  source catalog (\cite{hartman}). (Note: After submission of this paper, the EGRET group, using analysis based on the third EGRET 
  catalog, confirmed the finding of a $4.3\sigma$ significance source in the 30-100 MeV band, consistent in
  position with PSR B1509-58; D.J. Thompson, private communication).
  
  We performed a similar imaging study in the 100-300 MeV energy window and found also a $\sim
  2\sigma$ source feature near PSR B1509-58 above of the galactic and extra-galactic background, as well as the previously reported 
  EGRET sources, consistent with the findings of \cite{brazier1} and \cite{fierro}. 
  The 100-300 MeV flux estimate for PSR B1509-58 is $(5.9\pm3.5)\cdot 10^{-10}\ ph/cm^2\cdot s \cdot MeV$. 
  Even in the 300-1000 MeV window a marginal excess is visible consistent in position with PSR B1509-58 (flux 
  $(4.5\pm3.6)\cdot 10^{-11}\ ph/cm^2\cdot s \cdot MeV$). Above 1000 MeV no excess is found. These
  low flux values can also be converted in upper-limits.

  \begin{figure}[t]
     
     \vspace{-0.8truecm}
     \hbox{\hspace{0.50truecm}
           \psfig{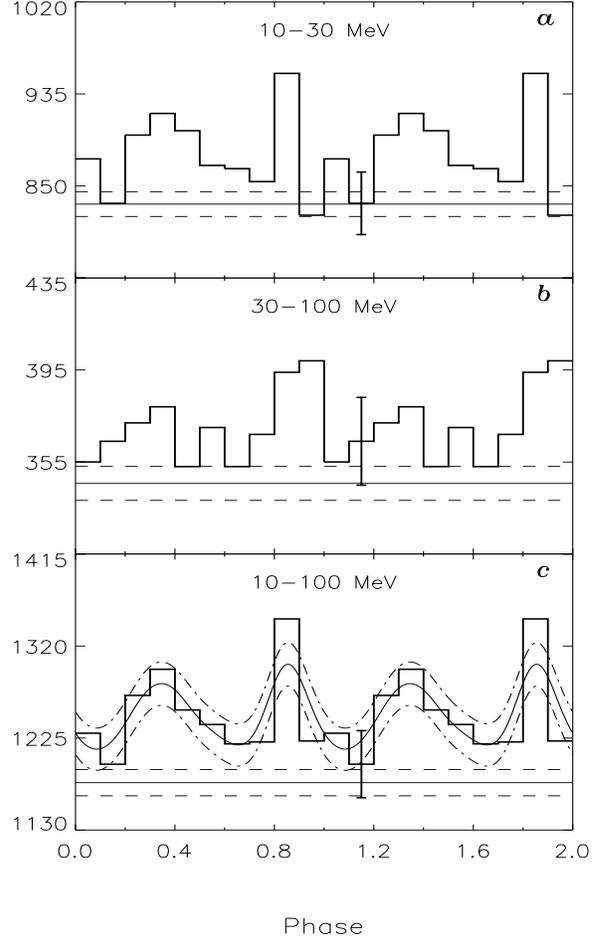}
          }
     \vspace{0.35truecm}
     \caption[]{{\bf a} COMPTEL 10-30 MeV Cycle I-VI lightcurve ($2.1\sigma$): the background level and its $\pm 
                        1\sigma$ error estimated from the spatial analysis are indicated as a solid straight line and 
                        dashed lines.  
                        {\bf b} EGRET 30-100 MeV Cycle I-IV lightcurve ($1.1\sigma$): background level (and $\pm  
                        1\sigma$ errors) are determined in a spatial analysis including point sources at the 
                        pulsar and 2EG J1443-6040 positions. {\bf c} Combined COMPTEL 10-30 MeV and EGRET 30-100 MeV
                        lightcurve ($2.3\sigma$ or 1.9\% chance probability). The summed background level from the spatial analyses
                        is again indicated together with the smoothed curves representing the KDE and its $\pm 1\sigma$ error
                        region. Typical error bars are indicated.}
     \label{fig:comptel_egret}
  \end{figure} 

  \subsection{Timing analysis}

  Given the detection of a significant source in the spatial analysis for energies between 30 and 100 MeV, we first selected events 
  in this energy range to search for the modulation. The event selection procedures in the EGRET timing analysis are equivalent to 
  those applied in the COMPTEL timing analysis except that we are now dealing with a 2-dimensional dataspace. 
  We applied an energy dependent cone selection (\cite{thompson3}) roughly selecting $68\%$ of the source counts.
  In this 30-100 MeV energy interval we do not find significant modulation ($1.1\sigma$), although the shape of the pulse phase 
  distribution bears some resemblance to the COMPTEL 10-30 MeV lightcurve. Both are shown in Fig. \ref{fig:comptel_egret}. 
  Estimates of the underlying backgrounds as determined from the spatial analysis fitting PSR B1509-58 and 2EG J1443-6040
  simultaneously are indicated. Also shown in the figure is the combined 10-100 MeV lightcurve, which has a modulation significance of 
  $2.3\sigma$ for 2 harmonics in the $Z_n^2$-test. Again, the summed background level determined in the spatial analyses is indicated, 
  as well as the Kernel Density Estimator (KDE, \cite{dejager}) of the unbinned pulse-phase distribution with the 
  $\pm 1\sigma$ error bounds. The KDE provides an (asymptotically) unbiased view on the genuine pulse shape.
  This lightcurve suggests a double-peaked profile: an enhancement in the previously defined ``pulsed'' interval, and a pulse 
  near phase 0.85, which has become more pronounced in this summed lightcurve. 
 
  Even though the modulation significance of this 10-100 MeV lightcurve is still marginally significant, the apparent double-peaked 
  structure above the background level estimated in the spatial analysis, makes it very suggestive that the source detected in the 
  spatial analysis is PSR B1509-58 with pulsed emission upto the EGRET energies. At least, the spatial and timing analyses are 
  consistent with this interpretation. If correct, the lightcurve morphology changed from one broad single pulse to a profile with 
  an additional pulse near phase 0.85.

%%%%%%%%%%%%%%%%%%%%%%%%%%%%%%%%%%%%%%%%%%%%%%%%%%%%%%%%%%%%%%%%%%%%%%%%%%%%%%%%

%%%%%%%%%%%%%%%%%%%%%%%%%%%%%%%%%%%%%%%%%%%%%%%%%%%%%%%%%%%%%%%%%%%%%%%%%%%%%%%%%%%%%%%%%%%%%%%%%%%%%%%%

\begin{figure}[t]
     \hbox{
           \hspace{-0.35cm} 
           \psfig{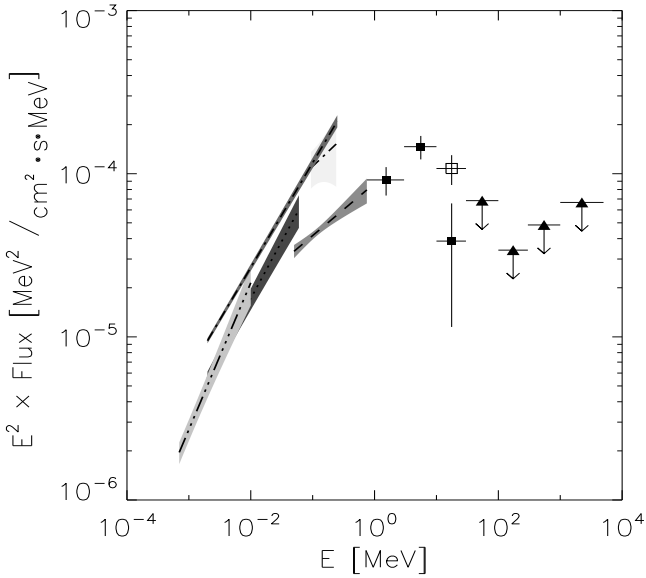}
          }
     \caption[]{The pulsed high-energy spectrum of PSR B1509-58 from soft X-rays to hard 
               $\gamma$-rays. The filled squares are the COMPTEL flux points as derived from the 
               excess counts in the $0.15-0.65$ phase range (case {\em i\/}), from section 3.5), while 
               the open square represents the 10-30 MeV flux in the 0.15-0.65 phase interval above the 
               spatially determined background (case {\em ii\/}). 
               The filled triangles are the $2\sigma$ upper limits for the total fluxes in the EGRET energy 
               domain (D. Thompson - private communication). 
               The polygons represents the best fit $1\sigma$ error regions as measured by different 
               instruments below the COMPTEL energy window: ASCA (0.7-10 keV; \cite{saito}); 
               Ginga (2-60 keV; \cite{kawai1}); OSSE (50-750 keV; Matz et al. 1994); WELCOME (94-240 keV; 
               Gunji et al. 1994); RXTE (2-250 keV; Marsden et al. 1998). The softening of the spectrum 
               from soft X-rays to MeV gamma rays is evident, as well as the spectral break above 10 MeV.
               }
     \label{fig:he_spectrum_timing}
\end{figure}

%%%%%%%%%%%%%%%%%%%%%%%%%%%%%%%%%%%%%%%%%%%%%%%%%%%%%%%%%%%%%%%%%%%%%%%%%%%%%%%%%%%%%%%%%%%%%%%%%%%%%%%%

\section{Spectral analysis}

Energy spectra can be derived from the spatial analysis as well as from the timing analysis. Given that
there is some ambiguity in the interpretation of the results on PSR B1509-58 for energies above 10 MeV, 
we consider three cases:

{\em i\/}) The COMPTEL ``pulsed'' spectrum of PSR B1509-58 (0.75-30 MeV) determined by the excess counts in 
the phase window 0.15-0.65 above the average level in the complementary phase interval. This represents the 
spectrum of the broad single pulse, which was already determined in section 3.5 (Table 3). In Fig. 
\ref{fig:he_spectrum_timing} this spectrum is shown together with the spectra (power-law fits) derived by 
other instruments from 1 keV upto $\sim$ 700 keV. 
Above 30 MeV are included the EGRET $2\sigma$ upper-limits (D.Thompson-private communication) derived 
from spatial analyses using a subset of the observations listed in Table 1. The new COMPTEL flux values
are consistent with an extrapolation of the OSSE power-law fit (PL-index of -1.68) upto 10 MeV, followed 
by a drastic break, which will be discussed in section 7.

{\em ii\/}) If the source detected by COMPTEL between 10 and 30 MeV is PSR B1509-58, then we have underestimated the 
flux of the single broad pulse in the 10 - 30 MeV interval. Determining then the flux (phases 0.15-0.65) above the 
spatially determined background (see Fig. \ref{fig:comptel_egret}) between 10 and 30 MeV, and inserting that in 
Fig. \ref{fig:he_spectrum_timing} gives a spectrum for the single broad pulse with a break energy between 10 and 30 MeV. 
In this case this spectrum does not represent the total pulsed spectrum for energies above 10 MeV.

\begin{figure*}[t]
     \hbox{ 
     \hspace{-0.125cm} 
     \centerline{\psfig{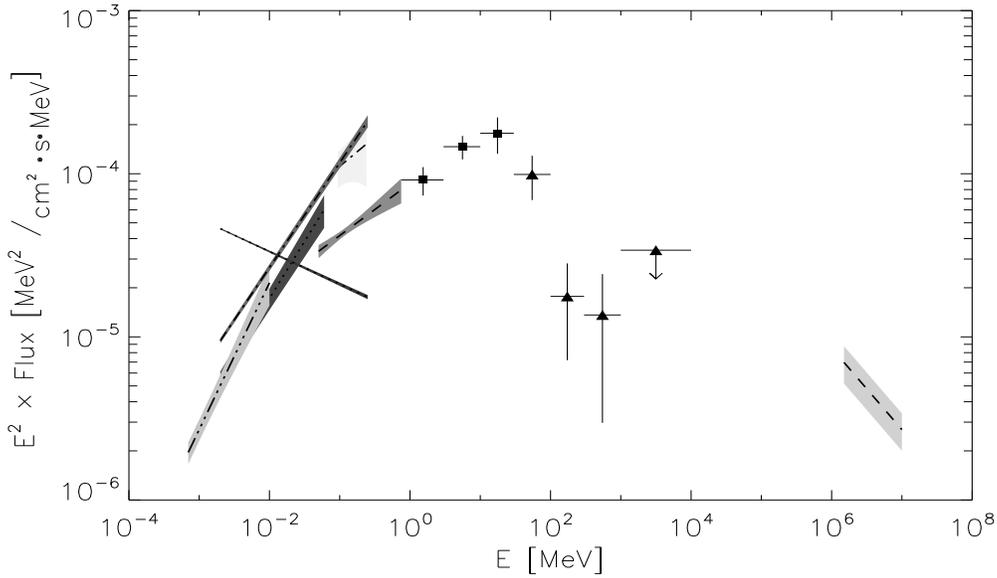}}
          }
                \caption[]{The {\em total} COMPTEL and EGRET high-energy spectrum of PSR B1509-58 ($>0.75$ MeV, 
                case {\em iii\/}) including the spectral measurements below 0.75 MeV of the pulsed emission by
                other high-energy experiments (see caption Fig. \ref{fig:he_spectrum_timing}).
                Furthermore, the ``unpulsed'' spectrum of PSR B1509-58 measured by RXTE (Marsden et al. 1998) is
                shown, as well as the $>1.5$ TeV DC-flux measurement by the CANGAROO collaboration (assumed to have
                a photon index of -2.5).
                The {\em total} emission breaks now around 30 MeV due to the contribution from an 
                additional pulse near phase 0.85 for energies in excess of 10 MeV.
               }
     \label{fig:he_spectrum_spatial}
\end{figure*}

{\em iii\/}) If the sources detected at the position of the pulsar by COMPTEL and EGRET in the spatial analyses 
above 10 MeV and 30 MeV, respectively, can be identified with the pulsar, then the {\em total\/} spectrum of 
PSR B1509-58 above 0.75 MeV can be derived by combining the COMPTEL flux values from the timing analysis below 10 MeV
(consistent with 100\% pulsed) with the flux values from the spatial analysis of COMPTEL 10-30 MeV data (section 4.1.2) and 
EGRET data above 30 MeV (section 5.1). This is shown in Fig. \ref{fig:he_spectrum_spatial} together with the ``pulsed'' spectra 
measured at lower energies by other instruments as well as the flux estimate from the detection ($4.1\sigma$) by the 
CANGAROO collaboration in their spatial analysis for energies above 1.5 TeV (\cite{sako}).
The increased flux level above 10 MeV in comparison with the spectrum shown in Fig. \ref{fig:he_spectrum_timing} is then 
mainly due to an additional pulse component near phase 0.85, having a completely 
different spectrum than the main pulse. For comparison is also shown the ``unpulsed'' spectrum of PSR B1509-58
measured by RXTE upto 240 keV (Marsden et al. 1998). It is evident that, if the ``unpulsed'' emission extrapolates according to the 
RXTE spectrum towards the $\gamma$-ray regime, the 10-30 MeV flux measured at the PSR B1509-58 position can {\em not\/} be the
(nebula) DC emission, supporting the pulsar interpretation. Furthermore, the EGRET flux values above 100 MeV are consistent 
with the extrapolation of the fit to this ``unpulsed'' spectrum, which is even consistent with an extrapolation up to the 
claimed detection of the nebula by the CANGAROO collaboration at TeV energies. Therefore, the spectral compilation
in Fig. \ref{fig:he_spectrum_spatial} suggests that the combined COMPTEL / EGRET spectrum of PSR B1509-58 represents the transition
from dominantly pulsed emission below $\sim 30$ MeV to dominantly DC-emission above $\sim 100$ MeV. 

%%%%%%%%%%%%%%%%%%%%%%%%%%%%%%%%%%%%%%%%%%%%%%%%%%%%%%%%%%%%%%%%%%%%%%%%%%%%%%%%%%%%%%%%%%%%%%%%%%%%%%%%
 
%%%%%%%%%%%%%%%%%%%%%%%%%%%%%%%%%%%%%%%%%%%%%%%%%%%%%%%%%%%%%%%%%%%%%%%%%%%%%%%%

  \section{Summary and discussion}

  The major findings based on primarily COMPTEL and to a lesser extent EGRET $\gamma$-ray data presented in this paper 
  can be summarized as follows:

  \begin{itemize}
    \item[\small{\bf{I}}]{Pulsed $\gamma$-ray emission from PSR B1509-58 has been detected with high significance up to 
                   10 MeV as a broad asymmetric single pulse 
                   located in the phase interval where also the pulse occurs for X-ray and soft $\gamma$-ray energies up to 
                   $\sim$ 700 keV. The pulse measured by COMPTEL between 0.75 and 10 MeV
                   reaches its maximum near radio phase 0.38, shifted w.r.t.  the value 0.30 measured by Rots et 
                   al. (1998) in the BATSE data above 32 keV. The COMPTEL maximum appears to coincide in phase wih the 
                   ``shoulder'' clearly visible in the RXTE 2-16 keV data. The pulsed spectrum is consistent with a power-law 
                   extrapolation up to 10 MeV of the OSSE spectrum (PL-index of -1.68) measured at lower $\gamma$-ray energies.}
    \item[\small{\bf{II}}]{In the 10-30 MeV energy range we do detect a significant ($5.6\sigma$) source positionally consistent
                     with PSR B1509-58, however, the timing analysis yields a modulation significance
                     of $\sim 2\sigma$ only. Based on just the COMPTEL data we cannot discriminate with certainty between
                     the following interpretations: {\em i\/}) DC emission from the pulsar or its nebula, {\em ii\/}) emission 
                     from a nearby source, or {\em iii\/}) pulsed emission: the light curve exhibits an excess in phase with the 
                     pulse measured below 10 MeV and there is an indication for a second narrow component near phase 0.85.} 
    \item[\small{\bf{III}}]{In the adjacent 30-100 MeV EGRET energy band we detect a $6.7\sigma$ source positionally
                    consistent with PSR B1509-58 which can be explained with contributions from PSR B1509-58 ($4.4\sigma$) and the
                    nearby unidentified EGRET source 2EG J1443-6040 ($2.6\sigma$). Timing analysis of the 30-100 MeV events yields a 
                    modulation significance of $1.1\sigma$ only. The combined COMPTEL/EGRET 10-100 MeV lightcurve (modulation
                    significance $2.3\sigma$) shows also the double-peaked shape: again a main pulse coinciding with the 
                    pulse observed at lower energies and a weak narrower pulse near phase 0.85. The results of the spatial and 
                    timing analyses are consistent with PSR B1509-58 being detected between 10 and 100 MeV, with a new pulse 
                    which is only visible between 10 and 100 MeV.}
    \item[\bf{IV}]{The pulsed spectrum of PSR B1509-58 shows a sharp break between 10 and 30 MeV. The break energy is close to 10 MeV
                    for the broad main pulse, and shifts to $\sim 30$ MeV for the total spectrum if the second narrow component is
                    genuine.}
  \end{itemize}

  The implications of these new findings at medium and high-energy $\gamma$-ray energies will now be discussed.
  Theoretical models explaining the high-energy electro-magnetic radiation from highly magnetized rotating neutron 
  stars can be distinguished in two different catagories: 
   
  \begin{itemize}
    \item[\bf{a}]{Polar Cap models (PC) in which the acceleration of charged particles along the open magnetic field lines 
                  in the vicinity of the magnetic pole(s) and subsequent cascades through high-energy radiation 
                  processes gives rise to the emerging $\gamma$-ray spectrum.}
    \item[\bf{b}]{Outer Gap models (OG) in which the acceleration of charged particles and the production of $\gamma$-rays 
                  takes place in vacuum gaps between the null-charge surface defined by $\bf \Omega\cdot B = 0$, with 
                  $\bf B$ the local magnetic field and $\bf \Omega$ the pulsar spin vector, and the light cylinder
                  ($R_{lc}=c/\Omega$) along the last closed field lines.}
  \end{itemize}

  Both models rely on charge replenishment of the magnetosphere mainly through $e^\pm$ pairs.
  The polar cap models can be subdivided further primarily based on the energy reached by the particles
  (primaries) in the acceleration process.
  For primary particle energies $\Gamma \ga 10^6$, with $\Gamma$ the Lorentz factor or equivalently the dimensionless 
  energy, energy loss in the form of high-energy $\gamma$-rays through {\em curvature radiation} is the most important 
  radiation mechanism (CRPC models). 
  For energies $10^4 \la \Gamma \la 10^6$ energy loss through {\em inverse Compton scattering} with 
  either thermal X-ray photons from the polar cap or from non-thermal cascade processes in the magnetosphere will be more 
  important (ICPC-models). 
  Finally, for energies $50 \la \Gamma \la 10^4$ {\em resonant Compton scattering} will be the dominant energy loss 
  mechanism.
  Irrespective of the underlying energy loss mechanism of the primary particles it is assumed that the emergent $\gamma$-ray
  spectrum is softened by magnetic pair production ($\gamma\stackrel{B}{\rightarrow}e^+ + e^-$; Sturrock process) in the 
  strong magnetic 
  fields present in the vicinity of the magnetic poles. These pairs (secondaries) are produced in excited Landau states, 
  which decay by the emission of synchrotron photons which in turn can produce $e^\pm$ pairs and a cascade can develop 
  softening the initial $\gamma$-ray spectrum at each generation. 

  Daugherty \& Harding (1996) assume in a recent version of their CRPC model that a distant observer sees the emission from just one 
  polar cap. They also assume that the acceleration of the primaries occurs over an extended 
  distance above the polar cap surface and that the magnetic inclination $\alpha$ (angle between $\bf \Omega$ and magnetic 
  moment {\boldmath $\mu$}) is comparable to the $\gamma$-ray beam opening angle $\theta_{\gamma}$. The last two
  assumptions relax the requirement of both $\alpha \approx \theta_{\gamma}$ and a small $\alpha$, i.e. a nearly aligned
  rotator, in earlier versions of their model (\cite{daugherty1}) to explain the $\gamma$-ray emission properties of
  pulsars.

  Because the curvature radii $R_{cr}$ of the open field lines, assuming a dipole configuration, originating from the 
  polar cap rim near $\theta_{pc} \simeq \arcsin(\sqrt(R_{ns}/R_{lc}))$ are smaller than those originating at smaller 
  polar angles (at the magnetic pole $R_{cr} \rightarrow \infty$ and no curvature emission can be produced)
  the initial CR $\gamma$-ray emission spectrum ($\propto 1/ R_{cr}$) is most pronounced and hardest (maximum of CR-spectrum 
  is reached at $\omega_{\max} \simeq 0.29 \cdot \omega_c \propto \Gamma^3 / R_{cr}$) 
  near the polar rim resulting in a hollow cone emission pattern.
  The mostly developed cascades originate near the polar rim which in turn soften the input CR-spectrum most efficiently.

  This means that a distant observer not only sees a double peaked profile when the hollow $\gamma$-ray emission
  cone passes his line of sight, but also a soft-hard-soft variation of the emission is expected, because the CR-emission 
  emanating nearer the magnetic pole is less softened by cascading. Grazing through the $\gamma$-ray emission cone will result 
  in a single pulse profile highly softened by cascade processes. The last situation may be applicable for PSR B1509-58  
  requiring clear constraints on the geometrical aspects of both the system and the observer. In particular, it demands that 
  the pulsar spin axis / line-of-sight angle $\zeta$ should approximately be equal to the sum of the magnetic inclination 
  angle $\alpha$ and the $\gamma$-ray cone semi-angle $\theta_{\gamma}$ (beam radius). However, the large duty cycle of the
  single pulse of $\sim 0.4$ poses an additional constraint, namely $\zeta$ should then be smaller than $\alpha+\theta_{\gamma}$.
  This can be estimated using the following equation:

  \begin{equation}
%      \sin^2({{\theta_{\gamma}}\over {2}})=\sin^2({{\Delta\Phi}\over{4}}) \sin\alpha \sin(\alpha+\beta) + \sin^2({{\beta}\over{2}})
      \theta_{\gamma} = \arccos(\cos\beta-2\sin\alpha\sin(\alpha+\beta) \sin^2({{\Delta\Phi}\over{4}}))
      \label{eq:gamma_beam}
  \end{equation}

  \noindent which can be derived in the rotating vector model (e.g. Rankin 1993). $\Delta\Phi$ is the duty cycle of the pulse, while 
  $\beta$ is the impact parameter defined as $\beta\equiv\zeta-\alpha$. 
  Radio-polarization data yield in the context of this rotating vector model (Crawford et al. 1997;
  private communication 1999) $\alpha\simeq 18^{+18}_{-13}$ degree and $\beta\simeq 20^{+20}_{-20}$ degree. If we use the best fit
  parameters, then $\theta_{\gamma} \simeq 36\degr$. Eventhough the uncertainties are large, the emitting rim must be very broad
  to explain the broad single pulse. 
 
%  which can be consistent with radio-polarization data yielding in the context of the rotating vector model (Crawford et al. 1997; 
%  private communication 1998) $\zeta\simeq 64^{+20}_{-31}$ degree and $\alpha\simeq 31^{+9}_{-15}$ degree (errors are at the $1\sigma$ 
%  confidence level), requiring $\theta_{\gamma} \approx \alpha$. However, to model the measured broad profile 
%  with a duty cycle of $\sim 0.5$, the actual values for $\zeta$ and $\alpha$ should both be close to their
%  1-$\sigma$ lower bounds.
  
  The CRPC model also predicts a spectral cutoff in the $\gamma$-ray spectrum at several GeV. 
  Another interesting feature is that a lower cutoff energy is expected for those pulsars with a higher
  magnetic field, because the softening by cascade processes is more efficient in these cases. Some of these general 
  trends are indeed observed in some $\gamma$-ray pulsars (\cite{thompson4}).
  The current implementation of geometrical and physical aspects, however, predicts lightcurves with a high degree of
  symmetry. This symmetry is a problem for e.g. the Vela pulsar high-energy lightcurve (\cite{grenier}; \cite{kanbach}) which 
  shows rather asymmetric bridge emission. 

  For PSR B1509-58 this CRPC model {\em can not} explain the observed {\em low} spectral cutoff energy occuring between 10 and 30 MeV. 
  However, the (very) strong polar surface magnetic field strength of $\sim 3.1\times10^{13}$\ Gau\ss \ likely triggers 
  another more exotic attenuation process to be active in the vicinity of the magnetic pole, namely {\em photon splitting}
  $\gamma\stackrel{B}{\rightarrow}\gamma\prime + \gamma\prime\prime$, besides attenuation by magnetic pair production.
  \cite{harding} showed that the photon splitting attenuation lengths can be shorter than the attenuation lengths for
  pair production for magnetic field strenghts $B \ga 0.3 B_{cr}$ with $B_{cr} = m_e c^3 / e\hbar = 4.413\times 10^{13}$ G.
  This means that photon splitting acts as the dominant attenuation process for such strong magnetic fields and can suppress
  the emission of high-energy photons. Contrary to the attenuation by magnetic pair production, photon splitting has {\em no}
  energy threshold and can degrade the photon energy also below the magnetic pair production threshold 
  ($\omega\ge 2m_ec^2/\sin(\theta_{kB})$, $\theta_{kB}$ is the angle between photon propagation direction and local magnetic 
  field).
  Depending on the splitting mode, partial and full splitting cascades are addressed, \cite{harding} calculate the
  high-energy spectrum of PSR B1509-58 for various values of the model parameters $\theta_{kB}$ and $\theta$, the magnetic
  co-latitude angle, assuming that the initial high-energy photon emission originates from the neutron star surface in a polar rim 
  at co-latitude $\theta$. Our new medium-energy $\gamma$-ray data and in particular the spectrum of 
  the main pulse severely constrain the magnetic co-latitude of the emission rim, irrespective of the splitting mode. 
  A co-latitude of $\sim 2\degr$, close to the ``classical'' radius of the polar cap $\theta_{pc}$, appears to be required 
  in the model calculations to be consistent with the evident spectral break between 10 and 30 MeV in the combined 
  COMPTEL-EGRET spectrum. 

  No detailed model calculations have been performed for PSR B1509-58 in the inverse Compton induced Polar Cap cascade scenario 
  (\cite{sturner1}; \cite{sturner2} and \cite{sturner3}), although some interesting qualitative statements have been made.
  In particular, if $B > 10^{13}$ G and the neutron star surface temperature $T > 3\cdot 10^6$ K the Lorentz factors of the electrons
  are limited to $\la 10^3$, which might explain the low cutoff energy in the spectrum of PSR B1509-58 
  (\cite{sturner3}).
  However, this type of PC-model also suffers from predicting too symmetric lightcurves. The model requirement of a nearly aligned
  rotator ($\alpha \sim \theta_{\gamma} \la 5\degr$) can formally not be excluded for the estimates of the magnetic inclination 
  $\alpha\simeq 18^{+18}_{-13}$ degree (Crawford et al. 1997) and $\theta_{\gamma} \simeq 36^{+32}_{-30}$ degree.

  In outer gap scenarios (see e.g. \cite{chenga}, \cite{chengb} and \cite{ho}) it is believed that stable vacuum gaps (\cite{holloway}) 
  can be formed in the outer magnetosphere along the 
  boundary of the last closed field lines which extend from the null-charge surface to the light cylinder. Voltage drops of typically
  $\sim 10^{13}$ V can be obtained across the gaps and accelerate $e^{\pm}$, created either in the gap or flowing in from beyond 
  the light cylinder and from across the null-charge surface, to energies limited by curvature radiation and to a lesser extent 
  inverse Compton scattering (off the ambient bath of lower energy photons). A geometrical calculation of the high-energy 
  emission (beamed along the local magnetic field in the outer gap resulting in a ``fan'' beam) from an outer gap region was 
  successfull in reproducing qualitatively the observed $\gamma$-ray lightcurves of the known $\gamma$-ray pulsars (\cite{chiang}).
  In a recent paper \cite{romani} modelled the emergent high-energy emission based on curvature radiation reaction limited accelerated 
  charges
  in the outer magnetosphere. In his model the primary particles ($e^{\pm}$) emit curvature radiation with spectral cutoffs in the
  1-10 GeV range. In order to tap the potential drop a small fraction (proportional to the optical depth $\tau_{\gamma \gamma}$ and of
  the order of $\la 10^{-3}$-$10^{-2}$) of these curvature photons interact with low-energy photons,
  provided by thermal surface emission from the neutron star or produced in the gap itself by synchrotron emission processes, to produce
  $e^{\pm}$ pairs. The perpendicular momentum of the produced pairs with respect to the local magnetic field is emitted in the form of
  synchrotron radiation peaking in the 1-10 MeV range. Part of the produced synchrotron radiation can inverse Compton scatter (ICS) off 
  the primary particles to produce a TeV pulsed emission component (typically $\la 1\%$ of the pulsed GeV flux). The composite of the 
  various emission components constitute the high-energy pulsed spectrum. 
  Although no quantitative estimates are presented, \cite{romani} argued
  that for short-period high-magnetic field pulsars like PSR B1509-58 the synchrotron flux produced in the gap itself will dominate the
  thermal surface emission increasing $\tau_{\gamma \gamma}$ towards $\sim 1$ and thus significantly suppressing the GeV curvature 
  component. This results in a synchrotron type spectrum peaking at MeV energies.
  From a geometrical point of view, the derived large values for the magnetic inclination and line-of-sight / pulsar spin axis 
  angles seem to be naturally explained with an outer magnetospheric origin for the high-energy emission.

  The above considerations for both the polar cap and outer gap scenarios show that further model calculations are required to make 
  decisive statements on the production site and mechanisms of the high-energy pulsed radiation. In the small sample of detected 
  high-energy pulsars PSR B1509-58 represents clearly a special case. 
  This might now even be more so, given that in the above discussions we did not yet introduce the possible detection of an additional
  pulse in the lightcurve for energies above 10 MeV. This component must have a significantly harder spectrum than that of the broad
  main pulse. Such an additional narrow component may at first sight be more easily incorporated in an outer-gap scenario. 
  In this respect it may be of interest to refer to earlier claims of detection of multiple and variable components in the lightcurve 
  of PSR B1509-58 at TeV energies ($>1.5$ TeV), which have been interpreted in the frame work of the outer gap scenario (\cite{dejager1}, 
  \cite{nel}). However, the earlier mentioned TeV observations by the CANGAROO collaboration do {\em not} confirm these claims, and
  provide a significantly lower upper-limit to the pulsed emission.

  A long exposure of the source in the transition region of COMPTEL and EGRET is required to better study the change in morphology of
  the lightcurve. Additional COMPTEL observations of this source have been scheduled (CGRO Cycle 8) and are aimed to provide better 
  insights in the source characteristics especially above 10 MeV.

%%%%%%%%%%%%%%%%%%%%%%%%%%%%%%%%%%%%%%%%%%%%%%%%%%%%%%%%%%%%%%%%%%%%%%%%%%%%%%%%

\begin{acknowledgements}
The \comptel project is supported by NASA under contract 
NAS5-26645, by the Deutsche Agentur f\"ur Raumfahrtangelenheiten (DARA) under 
grant 50 QV90968 and by the Netherlands Organisation for Scientific Research (NWO).
AC research is sponsored by the CONACYT grant 4142-E. We thank Arnold Rots for 
providing the RXTE and BATSE lightcurves, Yoshitaka Saito for providing the 
ASCA spectral and timing results prior to publication and Froney Crawford \&
Vicky Kaspi for corresponding radio polarization results. Finally, we thank Dave
Thompson for discussions on the EGRET findings. 
\end{acknowledgements}

%%%%%%%%%%%%%%%%%%%%%%%%%%%%%%%%%%%%%%%%%%%%%%%%%%%%%%%%%%%%%%%%%%%%%%%%%%%%%%%%

%%%%%%%%%%%%%%%%%%%%%%%%%%%%%%%%%%%%%%%%%%%%%%%%%%%%%%%%%%%%%%%%%%%%%%%%%%%%%%%%

\end{document}